\documentstyle[12pt,psfig]{article}

\newcommand{\be}{\begin{equation}}
\newcommand{\ee}{\end{equation}}
\newcommand{\bea}{\begin{eqnarray}}
\newcommand{\eea}{\end{eqnarray}}

\begin{document}
\begin{titlepage}

%\flushright{To Appear: }
\flushright{IP/BBSR/2001-14 }

\vspace{1in}

\begin{center}
\Large
{\bf A Note on  Noncommutative String theory and its low energy limit }

\vspace{1in}

\normalsize

\large{ Shesansu Sekhar  Pal }\\
\em {e-mail: shesansu@iopb.res.in}

\normalsize
\vspace{.7in}

{\em Institute of Physics \\
Bhubaneswar - 751005, India }

\end{center}

\vspace{1in}

\baselineskip=24pt
\begin{abstract}
%\noindent .....
The noncommutative string theory is described by embedding open string theory 
in a constant second rank antisymmetric $B_{\mu\nu}$ field and the
noncommutative gauge theory is defined by a deformed  $\star$
product. As a check, study of  various scattering amplitudes in both
noncommutative string and noncommutative gauge theory 
confirm  that in the $\alpha^{'}\rightarrow 0$ limit, the
noncommutative string theoretic amplitude 
goes over to the noncommutative gauge theoretic amplitude, and the couplings are related as
$g_{YM}=G_0\sqrt{\frac{1}{2\alpha^{'}}}$. Furthermore we  show that
in this limit  there will not be any correction  to the gauge theoretic 
action because of absence of massive modes. We get sin/cos 
factors in the scattering amplitudes  depending on the odd/even  number of 
external photons.

\end{abstract}

\end{titlepage}

%\newpage

%\flushleft{\section{Introduction}}
\section{Introduction}
Recently there have been interesting developments which exhibit
intimate connections between the noncommutative string theory and the
noncommutative gauge theories. It is
well known that in the zero slope limit, $\alpha^{'}\rightarrow 0$, the
scattering amplitudes computed from the noncommutative string theory reproduces the
corresponding noncommutative field theoretic results. For example, in the case of open
bosonic string theory which contains massless spin one particle in its 
spectrum. When one computes the tree level scattering amplitude
involving the gauge bosons in noncommutative string theory and takes $\alpha^{'}\rightarrow 0$ limit then 
the resulting amplitude coincides with the tree level noncommutative field theoretic 
amplitude. Moreover, generalizations to noncommutative non-Abelian gauge theories can
be accomplished using the well known prescription of introducing
Chan-Paton factors.\\

          It has been shown that the end points of  open strings
attached to D-branes do not commute if there is a constant B-field
along the brane directions\cite{sw,vs,ch}. Therefore, it is natural to
explore the connections between string theories (with noncommuting
coordinates) and the noncommutative field theories. Indeed, Seiberg and Witten in their
seminal paper, have investigated relations between noncommutative gauge 
theories and the noncommutative string theory.\\

The purpose of this paper is to carry out some explicit calculations
of scattering amplitudes of open string states in noncommutative
string theory and then take
$\alpha^{'}\rightarrow 0$ limit. Then we compare these results with that of 
the noncommutative field theoretic ones. We may recall that the
non-commutativity parameter, $\theta$, defined below is related to the
B-field and therefore setting $B=0$ amounts to going over to
commutative theory. 
  
The non-commutativity is defined as
$[X^{\mu},X^{\nu}]=i\theta^{\mu\nu}$. 
One can derive a stringy uncertainty relation for the noncommutative
coordinates, starting from the noncommutative algebra, which is
\begin{equation}
\Delta X^{\mu}\Delta X^{\nu}\geq\frac{|\theta^{\mu\nu}|}{2}
\end{equation}
From this equation it is very easy to see that  the small change in $X^{\mu}$
 is related to the large change in $X^{\mu}$. But we know that 
$\Delta X^{\mu}\rightarrow 0$
corresponds to the UV divergence whereas $\Delta X^{\mu}\rightarrow \infty$ 
corresponds to the presence of IR divergence. It therefore implies that   
this  kind of commutation relation  gives rise to the UV/IR mixing,  
but to see it's appearance explicitly  one needs to go to loop level.

The presence of $B_{\mu\nu}$ field makes the open string to behave differently in two distinct ways from that of 
without this field. (1)The scattering amplitude depends on the ordering of 
vertex operators, (2)NCOS theory: theories where open strings decouple from 
closed strings \cite{sst,gmms,km} and there by  makes a large class
of massive open string modes stable. Apart from this UV/IR mixing 
\cite{mrs}, as mentioned above, 
the field theory associated to the string theory, in the low energy limit, is 
nonlocal,
 because the fields in the action are multiplied by a (deformed) $\star$ product.
The deformed product is defined as 
\begin{equation}
A\star B(x)=\exp(i\frac{1}{2}
\theta^{\mu\nu}\partial^{y}_{\mu}\partial^{z}_{\nu})A(y)B(z)|_{y=z=x}
\end{equation}
This implies the presence of (infinitely many) derivatives in the action, 
hence  the theory becomes non-local. 
We must mention that $\star$ product has no effect on the
kinetic energy term under integral \footnote{Which can be seen in
  momentum space easily.} .

We can classify the noncommutative theories in 3 ways and the classification 
will be characterized through  the non-commutativity parameter
$\theta^{\mu\nu}$ i.e. depending on the nonvanishing components along different directions\footnote{Throughout in our calculations we shall take 
 $\mu, \nu...$ as spatial directions as we shall be dealing with
 space/space non-commutativity only.}.
$(1)$ Space/Space noncommutative theories: Which is defined as  $ \theta^{\mu\nu}\neq 0 $ and the other components of it as  0, one gets these type of
theories by suspending the D-brane in a background magnetic field. Moreover,
one can show that this theory has low energy limit i.e. $\alpha^{'}\rightarrow$0 limit and as has been suggested in \cite{agm} that unitary field theories can be derived from string theory in a limit.
 Since we shall show that space/space noncommutative field theories are
 derived from string theory implies, according to the above argument,
that these theories are unitary. In passing we mention that in this
 background there exists various dualities that has been
mentioned in \cite{sw,ps}.
$(2)$ Space/time noncommutative theories: Which has  nonvanishing 
components only along the time and spatial directions i.e.  
$\theta^{0\mu}\neq 0$, one gets these kind of theories by suspending 
open strings in the background electric fields.
As is shown in \cite{gm} that the field theory associated to this type of 
string theory will become 
non-unitary which means that this type of string theory  does not have a 
well-defined low energy limit.
$(3)$Light-like theories: Which is defined as  $\theta^{0\mu}=-\theta^{1\mu}$,
 these theories 
are the results of embedding open strings in both electric and magnetic fields.
These theories have a low energy or field theoretic limit which inturn implies
that the low energy theory  is unitary\cite{agm}. We have summarized these 
properties below.\\

\begin{tabular}{|l|l|r|}       \hline\hline
\emph{Type} & \emph{Field theory} & \emph{String theory}  \\  \hline
Space/space Non-commutativity \\ $\theta^{\mu\nu} \neq 0$ &
 Nonlocal and Unitary  & Nonlocal and Unitary   \\ \hline
Space/time Non-commutativity \\ $\theta^{0\mu} \neq 0$  
& Nonlocal , Anti-unitary  & Nonlocal and Unitary \\ \hline
Light-like Non-commutativity \\ $\theta^{0\mu}=-\theta^{1\mu}$ &
 Nonlocal and Unitary  & Nonlocal and Unitary   \\ \hline\hline
\end{tabular}\\
\space

In this $\alpha^{'}\rightarrow 0$ limit of noncommutative string theory, as we shall see at tree level  all the
massive open string states decouple and the relevant degrees of
freedom are the only massless open string states. As has been shown in
\cite{sw} there exists a decoupling limit in which 
the string dynamics  can be decoupled from the gauge  theory degrees of freedom, where the gauge field theory lives on a noncommutative space \cite{tf}.

The plan of the paper is as follows: In section 2 we shall review the four point
gauge boson amplitude in noncommutative string theory, which is of the
same form as that of Type I superstring four point amplitude in
commutative theory, except the phase factors that is multiplied with
the kinematic factors in each sectors.
We shall derive the 3-point amplitude for the scattering of  open strings  in
noncommutative string theory, and the SL(2,R) invariance can be fixed by 
inserting  the vertex operators at  0, 1, $\infty $, and
shall show that in the $\alpha^{'}\rightarrow 0$ limit the
noncommutative string theory
goes over to the corresponding noncommutative field theory, through explicit calculations. 
In section 3, we also derive four point amplitudes for the case of one photon, 
two photon, three photon with three tachyon, two tachyon and one tachyon
respectively in noncommutative bosonic string theory. \\

In this study  we have demonstrated explicitly that in the
$\alpha^{'}\rightarrow 0$ limit noncommutative 
string  theory reduces to the noncommutative gauge theory and in the
said limit  no  corrections to the noncommutative gauge
theoretic action is  observed due to the absence of
massive string modes. In particular we have shown  the form of the kinematic
factor (K) in the case of the 4-point noncommutative amplitude involving gauge bosons  to be of the same form  as that  in the corresponding commutative theory with the desired properties.    
We also   gets sin/cos factors in the noncommutative scattering
amplitudes depending
on the number of external photons. From the study of  the
noncommutative scattering  amplitudes we conclude that the phases
arises depending on 
the ordering of the vertex operators but in the total noncommutative amplitude
the presence of  sin/cos factor will be determined by the
coefficient of $e^{ik.X}$ in the vertex operator, particularly the 3-point
amplitude of photon and tachyon leads to the appearances of  phases  in the
total amplitude as 'sin' and 'cos' times the kinematic factors
respectively. The couplings of  noncommutative string theory $G_s$ and
the noncommutative gauge theory
$g_{YM}$ are related as $g_{YM}=\frac{G_0}{\sqrt {2\alpha^{'}}}$, Where $G_0$ 
is related to $G_s$ as mentioned earlier,
 which is of  the same form as that of the commutative theory.

\section{Tree level amplitudes}
We shall show explicitly that  the noncommutative field theoretic
amplitudes can be reproduced by taking the  $\alpha{'}\rightarrow 0$
limit  of the noncommutative string theory amplitudes. Also, we shall 
relate the couplings  of both the theories by demanding the equality of the 
tree level amplitudes in this limit.

\subsection{4-point amplitude of massless gauge bosons at tree level in
noncommutative superstring theory}
\begin{figure}[tb]
\psfig{figure=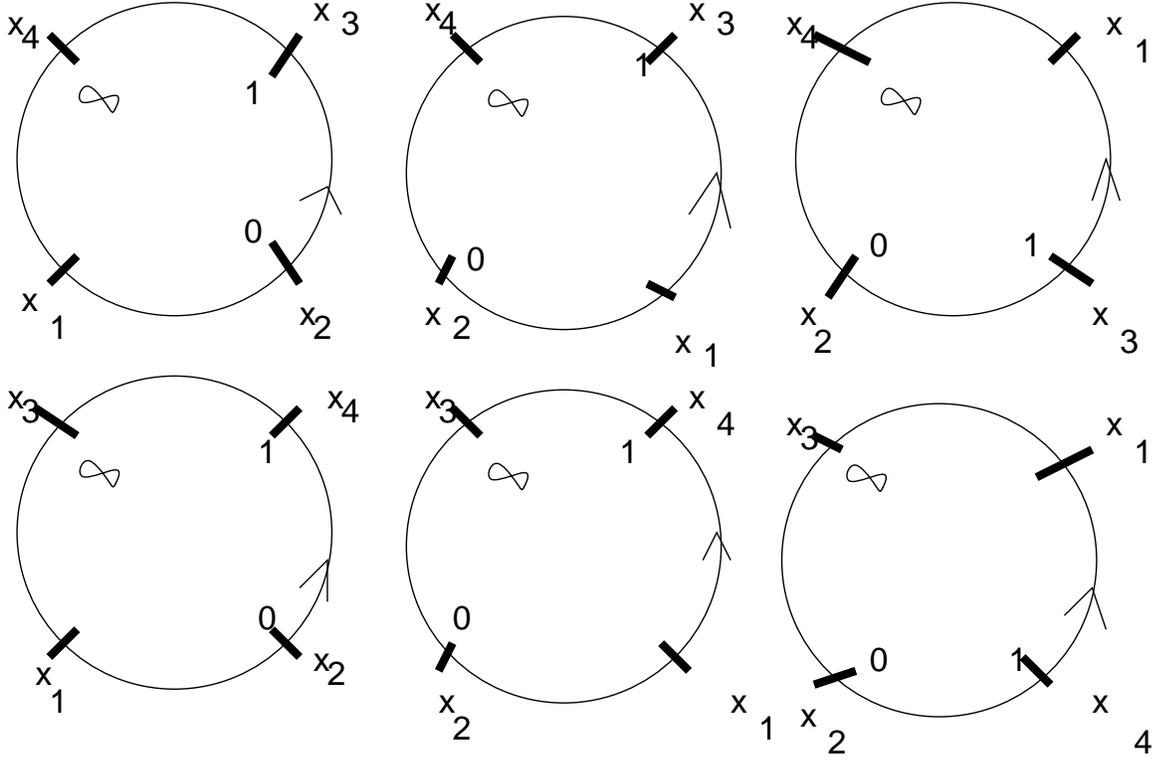,width=\hsize}
\caption{The vertex operators are inserted at $ x_1, x_2, x_3, x_4 $ on the
disk and we are fixing SL(2,R) by inserting  three of them at
0, 1, $\infty$. The position of the vertex operator increases along the
direction of arrow.}
\label{figb}
\end{figure}

In order to evaluate the scattering amplitude of four gauge bosons in
noncommutative string theory, we
shall follow the procedure of \cite{jp}. For the sake of completeness we
shall  demonstrate it. The scattering amplitudes are computed by
evaluating the correlation function of the vertex operators
corresponding to the asymptotic states in the scattering process. The form
of the amplitude for the processes that we are interested in is 

\begin{eqnarray}
A(e_1,p_1;e_2,p_2;e_3,p_3;e_4,p_4)&=&\int\frac{dx_{1}dx_{2}dx_{3}dx_{4}}{V_{CKG}}
\langle:e_{1}.V_{0}(p_{1},x_{1})::e_{2}.V_{0}(p_{2},x_{2}): \nonumber \\
& &:e_{3}.V_{-1}(p_{3},x_{3})::e_{4}.V_{-1}(p_{4},x_{4}): \rangle
\end{eqnarray}
Where $V_{CKG}$ is the volume of the  conformal killing group which remains 
as a residual gauge symmetry even after choosing the conformal gauge
and e's are the polarization vectors of the massless gauge bosons.
 The form of the vertex operators are
\begin{equation}
V^{\mu}_{0}(p,z)=G_0(2\alpha^{'})^{-\frac{1}{2}} (\partial
X^{\mu}(z)+2i\alpha^{'}p.\psi\psi^{\mu})e^{ip.X}
\end{equation}
\begin{equation} 
V^{\nu}_{-1}(p,z)=G_0e^{-\phi(z)}\psi^{\nu}(z)e^{ip.X}
\end{equation}
Where $G_0$  is defined as $G_0=16(\pi^{7/2})G_s(\alpha^{'2})$, $G_s$ is the open string coupling constant as defined in  \cite{sw}, which is $g_s(\frac{det(g+2\pi\alpha^{'}B)}{detg})^{\frac{1}{2}}$, g and B are the closed string fields . The subscript 0 and -1 in the vertex operator denote the super-ghost charge 
carried by the
vertex operators. The total ghost charge on a disk must be -2. This is a 
consequence of superdiffeomorphism invariance on the string world
sheet\cite{fms}. The following operator product expansion (OPE) will be 
useful for computation of scattering amplitudes
\begin{equation}
\langle X^{\mu}(x_1)X^{\nu}(x_2)\rangle=-\alpha{'}G^{\mu\nu}\log
(x_1-x_2)^2+\frac{i}{2}\theta^{\mu\nu}\epsilon (x_1-x_2)
\end{equation}
\begin{equation}
\langle
\centering{\psi^{\mu}(x_1)\psi^{\nu}(x_2)\rangle=-\frac{G^{\mu\nu}}{x_1-x_2}}
\end{equation}
$\langle \phi(x_1)\phi(x_2)\rangle=-\log(x_1-x_2)$\\
where $G^{\mu\nu}$ and $\theta^{\mu\nu}$ are as defined in \cite{sw}, which is 
$G^{\mu\nu}=(\frac{1}{g+2\pi\alpha^{'}B}g\frac{1}{g-2\pi\alpha^{'}B})^{\mu\nu}$ and $\theta^{\mu\nu}=-(2\pi\alpha^{'})^{2}(\frac{1}{g+2\pi\alpha^{'}B}B\frac{1}{g-2\pi\alpha^{'}B})^{\mu\nu}$.
We shall  map the disk on to the upper half plane and 
follow, what is called, the traditional method for
canceling the conformal Killing volume is to set the vertex operators 
on the real axis, as shown in the figure 1,  but there are 6 inequivalent 
ordering of the vertex operator on the disk. For the first 3 diagrams 
the integral of $x_1$ ranges from    $-\infty\leq
x_1 \leq 0$,  $0\leq x_1\leq 1$,   $1\leq x_1\leq \infty $ and the integration 
ranges for the last 3 diagrams can be obtained by 
 interchanging the vertex operator inserted at position $x_3$ with
$x_4$, i.e. $x_3\leftrightarrow x_4$. For the evaluation of the amplitude,
 we shall use transversality and masslessness
properties $e.p=0$ and $p.p=0$ respectively for the gauge bosons in our calculations.

In order to write the amplitudes in each sector in a manifestly Lorentz way, 
we shall 
introduce the well known Mandelstam variables for the gauge bosons
and their form is $s=-2 p_1.p_2$,  $t=-2 p_1.p_3$, and $u=-2 p_1.p_4$,  
since  $p^2_1=p^2_2=p^2_3=p^2_4=0$ for the gauge bosons. The amplitudes for the
first 3 diagrams are \footnote{$\Gamma(m)$ and $
B(m,n)=\frac{\Gamma(m)\Gamma(n)}{\Gamma(m+n)}$ are the well-known Gamma and Beta functions respectively.}

\begin{eqnarray}
A(s-t)=8\frac{G^2_{0}}{2\alpha^{'}}
         \frac{K}{st}\frac{\Gamma(1-\alpha^{'}s)\Gamma(1-\alpha^{'}t)}
          {\Gamma(1+\alpha^{' }u)} \times 
      e^{-i\frac{(p_1\theta p_2-p_3\theta p_4)}{2}} 
\end{eqnarray}
\begin{eqnarray}
A(t-u)=8\frac{G^2_{0}}{2\alpha^{'}}
          \frac{K}{ut}\frac{\Gamma(1-\alpha^{'}u)\Gamma(1-\alpha^{'}t)}
              {\Gamma(1+\alpha^{'}s)}\times 
      e^{i\frac{(p_1\theta p_4+p_2\theta p_3)}{2} }
\end{eqnarray}
\begin{eqnarray}
A(s-u)=8\frac{G^2_{0}}{2\alpha^{'}}
         \frac{K}{us}\frac{\Gamma(1-\alpha^{'}u)\Gamma(1-\alpha^{'}s)}
          {\Gamma(1+\alpha^{'}t)} \times
      e^{i\frac{(p_1\theta p_2+p_3\theta p_4)}{2}}
\end{eqnarray}
It is obvious to see that the phases of the amplitudes in each sector  are 
not derivable
by starting from any one of the amplitude, which we can say that the total  
amplitude in noncommutative string theory is not symmetric in the Mandelstam variables. The indices are
contracted as, $p\theta k=p_{\mu}\theta^{\mu\nu}k_{\nu}$
Where K is
\begin{eqnarray}
K&=&(2\pi)^{10}\delta^{10}(\sum_{i} p_i)\times[-\frac{1}{4}(ut e_1.e_2 e_3.e_4
                           +su  e_1.e_3 e_2.e_4+st e_1.e_4 e_2.e_3)\nonumber \\
 & &+\frac{s}{2}(e_2.p_4 e_1.e_4 e_3.p_1+e_2.e_4 e_1.p_4 e_3.p_2+e_2.p_3 e_4.p_1
                 e_1.e_3+e_1.p_3 e_4.p_2 e_2.e_3)+\nonumber  \\
 & &\frac{t}{2}(e_1.e_4 e_3.p_4 e_2.p_1+e_4.p_3 e_1.p_2 e_2.e_3+
                            e_2.p_3 e_1.p_4 e_3.e_4+e_1.e_2 e_3.p_2
                            e_4.p_1)+ \nonumber \\
 & &\frac{u}{2}(e_1.e_2 e_3.p_1 e_4.p_2+e_3.p_4 e_1.p_2 e_2.e_4+e_4.p_3 e_2.p_1 e_1.e_3+
                e_2.p_4 e_1.p_3 e_3.e_4)]
\end{eqnarray}
To evaluate the rest of the 3 diagrams interchange $x_3$ with $x_4$ and $p_3$ 
with $p_4$.
Thus the total amplitude for the scattering of 4 massless gauge bosons at
tree level of the noncommutative string theory in the $\alpha^{'}\rightarrow 0$ limit is
\begin{equation}
A^{Abelian}_{Total}=16\frac{G^2_{0}}{2\alpha^{'}}[\frac{K}{st}cos(\frac{p_1\theta p_2-p_3\theta p_4}{2})+
                       \frac{K}{su}cos(\frac{p_1\theta p_2+p_3\theta p_4}{2})+
                        \frac{K}{ut}cos(\frac{p_1\theta p_4+p_2\theta p_3}{2})]
\end{equation}
It is easy to derive the non-Abelian noncommutative string theory amplitude by multiplying the trace of
the product of the Chan-Paton factors to the amplitudes in different
sectors in accordance with the insertion of the  vertex operators. Finally the
full (non-Abelian noncommutative) 4-point amplitude is 
\begin{equation}
[A(s-t)tr(\lambda^{a_2}\lambda^{a_1}\lambda^{a_3}\lambda^{a_4})+
A(u-t)tr(\lambda^{a_2}\lambda^{a_3}\lambda^{a_1}\lambda^{a_4})+
A(s-u)tr(\lambda^{a_1}\lambda^{a_2}\lambda^{a_3}\lambda^{a_4})+3\leftrightarrow 4]
\end{equation}

In the $ \theta \rightarrow 0 $ limit this amplitude matches with  the 
four point amplitude of Type I superstring theory\cite{js}. 
The scattering amplitudes of massless scalars can be derived by
choosing the polarizations to lie in the transverse direction,
$e_l.p_m=0$. For example, the four massless scalar amplitude in
noncommutative string theory in the $\alpha^{'}\rightarrow 0$ limit is

\begin{eqnarray}
A_{scalar}&=&16\frac{G^2_{0}}{2\alpha^{'}}[\frac{M}{st}cos(\frac{p_1\theta p_2-p_3\theta p_4}{2})+
                         \frac{M}{su}cos(\frac{p_1\theta p_2+p_3\theta p_4}{2})+
                        \frac{M}{ut}cos(\frac{p_1\theta p_4+p_2\theta p_3}{2})] \nonumber \\
M&=&(2\pi)^{10}\delta^{10}(\sum_{i} p_i)\times[-\frac{1}{4}(ut e_1.e_2 e_3.e_4
                           +su  e_1.e_3 e_2.e_4+st e_1.e_4 e_2.e_3)
\end{eqnarray}

From the expression $2.12$ one can derive the scattering amplitude for 
two gauge bosons with two scalars, three gauge bosons with one scalar
and one gauge boson with three scalars using $e_l.p_m=0$ in the kinematic
factor K.\\

The intermediate particles that  propagates in the s-t channel of
the $U(1)$ noncommutative string theoretic amplitude are ,
$\alpha^{'}m^2=0,1,2,3,.....$, but going to it's low energy limit, one sees 
from 2.12 that only the massless particles propagates at the intermediate stages.
\subsection{Noncommutative gauge theoretic amplitude at tree level}
The action for the noncommutative gauge theory that we are interested in  is 
\cite{sw} 
 
\begin{eqnarray}
S=-\frac{1}{4}\int d^{10}x {\hat F_{\mu\nu}}\star {\hat F^{\mu\nu}} 
\end{eqnarray}
Where ${\hat F_{\mu\nu}}=\partial_{\mu}{\hat A_{\nu}}-\partial_{\nu}{\hat A_{\mu}}
-ig_{YM}[{\hat A_{\mu}},{\hat A_{\nu}}]_{\star}$, the fields are multiplied
through $ \star $ product and it is defined as $ A\star B(x)=\exp(i\frac{1}{2}
\theta^{\mu\nu}\partial^{y}_{\mu}\partial^{z}_{\nu})A(y)B(z)|_{y=z=x}. $
The Feynman rules for this action are
\begin{figure}[tb]
\psfig{figure=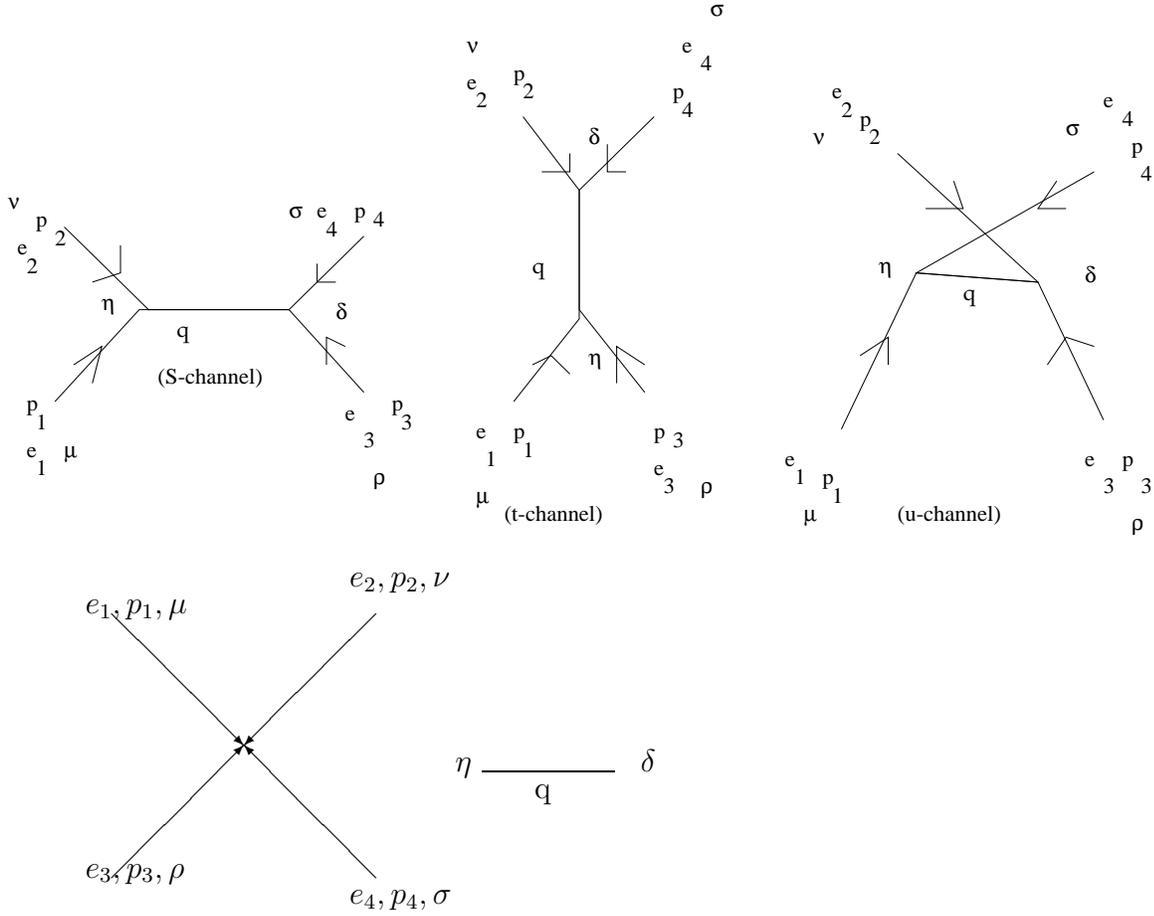,width=\hsize}
%\psfig{figure=fig2.ps,width=\hsize}
%\begin{figure}
%\begin{center}
\begin{picture}(200,200)(0,0)

%drawing the vectors
\put (40,70){\vector(1,1){50}}
\put(140,170){\vector(-1,-1){50}}
\put (40,170){\vector(1,-1){50}}
\put(140,70){\vector(-1,1){50}}
%writing the vertices
\put(30,70){$e_3,p_3,\rho$}
\put(130,180){$e_2,p_2,\nu$}
\put(30,170){$e_1,p_1,\mu$}
\put(130,60){$e_4,p_4,\sigma$}
%for propagator
\put(180,110){\line(1,0){50}}
\put(170,110){$\eta$}
\put(240,110){$\delta$}
\put(200,100){q}
\end{picture}
%\end{center}
\caption{Feynman diagrams for the noncommutative gauge theory at tree level.}
\label{figb}
\end{figure}
\begin{eqnarray}
V(p_1;p_2;p_3)&=&2ig_{YM}(2\pi)^{10}\delta^{10}(p_1+p_2+p_3)
[G^{\mu\rho}(p_3-p_1)^{\nu}+G^{\rho\nu}(p_2-p_3)^{\mu}+\nonumber \\
& & G^{\nu\mu}(p_1-p_2)^{\rho}]sin(\frac{p_1\theta p_2}{2}) 
\end{eqnarray}
%\newpage
\begin{eqnarray}
V(p_1;p_2;p_3;p_4)&=&4g^2_{YM}(2\pi^{10})\delta^{10}(p_1+p_2+p_3+p_4)
[sin(\frac{p_1\theta p_2}{2})sin(\frac{p_3\theta p_4 }{2}) \nonumber \\
& &(G^{\mu\rho}G^{\nu\sigma}-G^{\mu\sigma}G^{\nu\rho})+ 
sin(\frac{p_1\theta p_3}{2})sin(\frac{p_2\theta p_4 }{2})
(G^{\mu\nu}G^{\rho\sigma}-G^{\mu\sigma}G^{\nu\rho})+ \nonumber \\
& &sin(\frac{p_1\theta p_4}{2})sin(\frac{p_2\theta p_3 }{2})
(G^{\mu\nu}G^{\rho\sigma}-G^{\mu\rho}G^{\nu\sigma})]
\end{eqnarray}

We shall use the    propagator as  $-\frac{G^{\eta\delta}}{q^2}$ in
the evaluation of the noncommutative amplitude.  
The corresponding diagrams are shown in figure 1. $V(p_1;p_2;p_3)$ and $V(p_1;p_2;p_3;p_4)$
are the 3-point and the 4-point vertex functions
respectively. So at tree level i.e. to order $g^2_{YM}$ there are
4 diagrams, one is the contact interaction term and the other 3 are s, t
and u channel diagrams. The total noncommutative amplitude to order $g^2_{YM}$ is \\
%$A_{Total}=$
\begin{eqnarray}
A_{Total}&=&16g^2_{YM}[\frac{K}{st}cos(\frac{p_1\theta p_2-p_3\theta
  p_4}{2})+\frac{K}{su}cos(\frac{p_1\theta p_2+p_3\theta p_4}{2})+\nonumber \\
& &\frac{K}{ut}cos(\frac{p_1\theta p_4+p_2\theta p_3}{2})]
\end{eqnarray}
The form of this noncommutative gauge theoretic amplitude is same as that of the
noncommutative string theoretic amplitude 2.12 apart from over all 
constants. This implies  that there will not be any correction to the
noncommutative gauge
theoretic action  due to the absence of other modes. Now 
comparing the amplitudes of  noncommutative string theory with
noncommutative  gauge theory, we get a 
relation between the string parameters with the Yang-Mills coupling
namely, $g^2_{YM}(2\alpha^{'})=G^2_0$\\

\subsection{3-point amplitude in noncommutative open string theory} 
The 3-point noncommutative scattering amplitude of open string massless modes on the disk is
\begin{equation}
A(e_1,p_1;e_2,p_2;e_3,p_3)=\frac{1}{\alpha^{'}G^2_0}\langle
:cV_{-1}(p_1,x_1)::cV_{-1}(p_2,x_2)::cV_{0}(p_3,x_3):\rangle
+1\leftrightarrow 2
\end{equation}
The vertex operators are written as earlier and here c stands for  are
the ghost fields moreover  we need super ghost charge
on the disk to be -2 as mentioned earlier. Let us map the disk to the upper 
half plane and
fix the vertex operators at $x_3=0, x_2=1, x_1=\infty$ to cancel 
the residual SL(2,R) 
invariance that is left over even after working in the conformal gauge. We 
shall  evaluate the correlation function using the OPE's as 
written earlier and next, going over to  $\alpha^{'}\rightarrow 0$
limit we find that the noncommutative amplitude derived agrees with the one obtained from
the corresponding $U(1)$ noncommutative gauge theory apart from over all 
constants. The 3-point 
amplitude in $U(1)$ noncommutative  gauge theory is given by \\
%\begin{eqnarray}
$A(e_1,p_1;e_2,p_2;e_3,p_3)=2ig_{YM}(2\pi)^{10}\delta^{10}(\sum_i p_i)\times $
\begin{eqnarray}
[e_1.e_3 (p_3-p_1).e_2+ e_2.e_3(p_2-p_3).e_1 +e_1.e_2 (p_1-p_2)]sin(\frac{p_1\theta p_2}{2})
\end{eqnarray}
On looking at the kinematic factors we see that in the $\alpha^{'}\rightarrow 0$ limit there are not
any poles, which implies that the noncommutative gauge theoretic action will  remain 
unchanged, at least in tree level, as suggested in\cite{sw}.
The coupling in both the noncommutative theories are related as 
$g_{YM}=G_0\sqrt{\frac{1}{2\alpha^{'}}}$. This relation between the 
couplings are in the same form  as found in the
commutative theory. The corresponding non-Abelian noncommutative 3-point amplitude 
in the $\alpha^{'}\rightarrow 0$ limit  is same as that of the
noncommutative gauge theoretic one and it  is \\
\begin{eqnarray}
A&\sim&(2\pi)^{10}\delta^{10}(\sum_i p_i)[e_1.e_3
(p_3-p_1).e_2+e_2.e_3(p_2-p_3).e_1+e_1.e_2(p_1-p_2)]\times \nonumber \\
& &[e^{-i\frac{p_1\theta p_2}{2}}tr(\lambda^{a_1}\lambda^{a_2}\lambda^{a_3})-e^{i\frac{p_1\theta p_2}{2}}tr(\lambda^{a_2}\lambda^{a_1}\lambda^{a_3})]
\end{eqnarray}
\section{4-point amplitudes in noncommutative open bosonic string theory}
In this section we shall evaluate  4-point amplitudes at tree
level in noncommutative  bosonic string theory, and consider cases where the 
external particles are gauge bosons and tachyons. We shall show that if we are
taking odd number of photons then the amplitude will have a 'sin'
factor and for even number of photons it will have a cosine
factor. Before evaluating the scattering amplitude of photons with the
tachyons, we shall evaluate the 3-point and 4-point tachyonic amplitude which
will demonstrate that the 'sin' factor arises due to the presence of
odd number of photons (even though we are checking it by taking tachyons 
only).\\
{\large{\underline{3-point tachyonic amplitude:}}}\\
%\subsection{3-point tachyonic amplitude}
The noncommutative amplitude is 
\begin{equation}
A(p_1;p_2;p_3)=\frac{1}{\alpha^{'}G^2_{0}}x_{12}x_{13}x_{23} \langle
V(p_1)V(p_2)V(p_3)\rangle+2\leftrightarrow 3
\end{equation}
where $V(p)=G_{0}e^{ip.X}$, we shall evaluate it by fixing  the vertex
operators at  $x_1=1, x_2=0, x_3=\infty$ which will take care of the SL(2,R) 
invariance and using the mass shell condition as $p^2=\frac{1}{\alpha^{'}}$, we get  
\begin{equation}
A(p_1;p_2;p_3)=2\frac{G_0}{\alpha^{'}}(2\pi)^{26}\delta^{26}(\sum_i p_i) cos(\frac{p_1\theta p_2}{2})
\end{equation}
{\large{\underline{Scattering of  four tachyons:}}}\\
%\subsection{\Large{Scattering of  four tachyons}}
As has been mentioned in the evaluation of 4 massless gauge
bosons, there are 6 diagrams that one has to evaluate in order to derive 
the total amplitude, and  there are 2 diagrams in each channels.
The amplitude in the s-u channel can be derived by fixing the vertex 
operators at $x_2=0, x_3=1, x_4=\infty$ plus $3\leftrightarrow 4$, in order to 
cancel the SL(2,R) invariance and for these 2 diagrams the integration 
ranges of  $x_1$ are from $-\infty$ to 0 and 0 to 1 respectively. The 
form of the bosonic photon vertex that we are taking is
$V(e:p)=\frac{G_{0}}{\sqrt{\alpha^{'}}}e.\partial
Xe^{ip.X}$. The noncommutative amplitude in this sector is 
\begin{equation}
A_{su}(p_1;p_2;p_3;p_4)=-2\frac{G^2_0}{\alpha^{'}}(2\pi)^{26}\delta^{26}(\sum_i 
p_i)cos(\frac{p_1\theta p_2+p_3\theta
  p_4}{2})\frac{\Gamma(-1-\alpha^{'}s)\Gamma(-1-\alpha^{'}u)}{\Gamma(2+\alpha^{'}t)}
\end{equation}
The amplitude in the s-t sector can be derived by fixing the vertex
operators at $x_2=0, x_3=1, x_4=\infty$ and integrating $x_1$ from 0 to 1,
and for the 2nd diagram in this sector we fix the vertex operators at 
$x_2=0, x_3=\infty, x_4=1 $ and
integrating $x_1$ from $-\infty $ to 0. We get  the noncommutative amplitude in this sector as
\begin{equation}
A_{st}(p_1;p_2;p_3;p_4)=-2\frac{G^2_0}{\alpha^{'}}(2\pi)^{26}\delta^{26}(\sum_i 
p_i)cos(\frac{p_1\theta p_2-p_3\theta
  p_4}{2})\frac{\Gamma(-1-\alpha^{'}s)\Gamma(-1-\alpha^{'}t)}{\Gamma(2+\alpha^{'}u)}
\end{equation}
For the amplitude in the u-t channel, we fix the vertex
operators at  $x_2=0, x_3=1, x_4=\infty$ and $x_2=0, x_3=\infty, x_4=1 $ and 
integrating $x_1$ from 1 to $\infty$ for these two diagrams, one gets
the noncommutative amplitude in this sector as  
\begin{equation}
A_{ut}(p_1;p_2;p_3;p_4)=-2\frac{G^2_0}{\alpha^{'}}(2\pi)^{26}\delta^{26}(\sum_i 
p_i)cos(\frac{p_1\theta p_4+p_2\theta
  p_3}{2})\frac{\Gamma(-1-\alpha^{'}u)\Gamma(-1-\alpha^{'}t)}{\Gamma(2+\alpha^{'}s)}
\end{equation}
The total amplitude is the sum of the amplitudes in each sector.\\
{\large{\underline{Scattering of one photon with 3 tachyons:}}}\\
%\subsection{Scattering of one photon with 3 tachyons}
Here the fixing of the vertex operators shall be the same as in the
previous subsection and photon has been inserted at $x_2$. The total
noncommutative 
amplitude is the sum of amplitudes in each channel, $A_{st}+A_{su}+A_{ut}$,
 which is 
\begin{eqnarray}
& &A(p_1;e_2,p_2;p_3;p_4)=-4\frac{G^2_0}{\sqrt{ \alpha^{'}}} \times   
[sin(\frac{p_1\theta p_2+p_3\theta p_4}{2})(e_2.p_3(1+\alpha^{'}t)-e_2.p_4(1+\alpha^{'}u)) \nonumber \\
& &\frac{\Gamma(-1-\alpha^{'}s)\Gamma(-1-\alpha^{'}u)}{\Gamma(2+\alpha^{'}t)}- 
sin(\frac{p_1\theta p_2-p_3\theta p_4}{2})(e_2.p_3(1+\alpha^{'}t)-e_2.p_4(1+\alpha^{'}u))\nonumber \\
& &\frac{\Gamma(-1-\alpha^{'}s)\Gamma(-1-\alpha^{'}t)}{\Gamma(2+\alpha^{'}u)}- 
sin(\frac{p_1\theta p_3+p_2\theta p_4}{2})(e_2.p_3(1+\alpha^{'}t)-e_2.p_4(1+\alpha^{'}u))\nonumber \\
& &\frac{\Gamma(-1-\alpha^{'}t)\Gamma(-1-\alpha^{'}u)}{\Gamma(2+\alpha^{'}s)}]
\end{eqnarray}
{\large{\underline{Scattering of two photons with two tachyons:}}}\\
%\subsection{Scattering of two photons with two tachyons}
Here the  fixing of the vertex operators is same but the photons are
inserted  at $x_1, x_2$ and the tachyons are at $x_3$ and $x_4$. Total
noncommutative amplitude in this case is 
\begin{eqnarray}
&
&A(e_1,p_1;e_2,p_2,p_3,p_4)=\frac{4G^2_0}{\alpha^{'}}[cos(\frac{p_1\theta 
  p_4+p_2\theta p_3}{2})((e_1.e_2-2\alpha^{'}(e_1.p_3e_2.p_3+e_1.p_4e_2.p_4))\nonumber \\
&
&(1+\alpha^{'}t)(1+\alpha^{'}u)+2\alpha^{'}(e_2.p_4e_1.p_3\alpha^{'}u(1+\alpha^{'}u)+e_2.p_3e_1.p_4\alpha^{'}t(1+\alpha^{'}t)))\nonumber \\
&
&\frac{\Gamma(-1-\alpha^{'}u)\Gamma(-1-\alpha^{'}t)}{\Gamma(2+\alpha^{'}s)}+cos(\frac{p_1\theta
  p_2+p_3\theta
  p_4}{2})(-(e_1.e_2-2\alpha^{'}(e_1.p_3e_2.p_3+e_1.p_4e_2.p_4))\nonumber \\
&
&(1+\alpha^{'}t)(1+\alpha^{'}u)-2\alpha^{'}(e_2.p_4e_1.p_3\alpha^{'}u(1+\alpha^{'}u)+e_2.p_3e_1.p_4\alpha^{'}t(1+\alpha^{'}t)))\nonumber \\
&
&\frac{\Gamma(-1-\alpha^{'}u)\Gamma(-1-\alpha^{'}s)}{\Gamma(2+\alpha^{'}t)}+cos(\frac{p_1\theta 
  p_2-p_3\theta p_4}{2})(-(e_1.e_2-2\alpha^{'}(e_1.p_3e_2.p_3+e_1.p_4e_2.p_4))\nonumber \\
& &(1+\alpha^{'}t)(1+\alpha^{'}u)-2\alpha^{'}(e_2.p_4e_1.p_3\alpha^{'}u(1+\alpha^{'}u)+e_2.p_3e_1.p_4\alpha^{'}t(1+\alpha^{'}t)))\nonumber \\
& &\frac{\Gamma(-1-\alpha^{'}s)\Gamma(-1-\alpha^{'}t)}{\Gamma(2+\alpha^{'}u)}]
\end{eqnarray}
Instead of inserting the photons at   $x_1, x_2$ and the tachyons at
$x_3, x_4$, if we shall insert the tachyons at $x_2, x_4$ and photons at 
$x_1, x_3$ then the total noncommutative amplitude is 
\begin{eqnarray}
& &A(e_1,p_1;p_2;e_3,p_3;p_4)=\frac{4G^2_0}{\alpha^{'}}[cos(\frac{p_1\theta 
  p_2+p_3\theta p_4}{2})(e_1.e_3(1+\alpha^{'}u)(1+\alpha^{'}s)+\nonumber \\
&
&2\alpha^{'}(e_1.p_3e_3.p_1\alpha^{'}u(1+\alpha^{'}u)+e_1.p_4e_3.p_2\alpha^{'}t(1+\alpha^{'}t)-(e_1.p_3e_3.p_2+e_1.p_4e_3.p_1)\nonumber \\
&
&(1+\alpha^{'}u)(1+\alpha^{'}t)))\frac{\Gamma(-1-\alpha^{'}s)\Gamma(-1-\alpha^{'}u)}{\Gamma(2+\alpha^{'}t)}+cos(\frac{p_1\theta
  p_2-p_3\theta p_4}{2})\nonumber \\
& &(-e_1.e_3(1+\alpha^{'}u)(1+\alpha^{}s)+2\alpha^{'}(-e_1.p_3e_3.p_1\alpha^{'}u(1+\alpha^{'}u)-e_1.p_4e_3.p_2\alpha^{'}t(1+\alpha^{'}t)\nonumber \\
& &+(e_1.p_3e_3.p_2+e_1.p_4e_3.p_1)(1+\alpha^{'}u)(1+\alpha^{'}t)))\frac{\Gamma(-1-\alpha^{'}s)\Gamma(-1-\alpha^{'}t)}{\Gamma(2+\alpha^{'}u)}+\nonumber \\
& &cos(\frac{p_1\theta p_2+p_3\theta
  p_4}{2})(-e_1.e_3(1+\alpha^{'}u)(1+\alpha^{}s)+2\alpha^{'}(-e_1.p_3e_3.p_1\alpha^{'}u(1+\alpha^{'}u)-\nonumber \\
& &e_1.p_4e_3.p_2\alpha^{'}t(1+\alpha^{'}t)+(e_1.p_3e_3.p_2+e_1.p_4e_3.p_1)(1+\alpha^{'}u)(1+\alpha^{'}t)))\nonumber \\
& &\frac{\Gamma(-1-\alpha^{'}t)\Gamma(-1-\alpha^{'}u)}{\Gamma(2+\alpha^{'}s)}]
\end{eqnarray}
{\large{\underline{Scattering of 3-photons with one tachyons:}}}\\
%\subsection{Scattering of 3-photons with one tachyons}
Here vertex operators are fixed as in earlier cases, but the tachyon
is at $x_1$ and photons are located at rest positions. Total
noncommutative amplitude is
\begin{eqnarray}
&
&A(p_1;e_2,p_2;e_3,p_3;e_4,p_4=\frac{8G^2_0}{\sqrt{\alpha^{'}}}(sin(\frac{p_1\theta
  p_2+p_3\theta p_4}{2})[e_2.e_3e_4.p_1
B(-1-\alpha^{'}u,-\alpha^{'}s)+\nonumber \\
& &(e_2.e_3e_4.p_2-e_2.e_4e_3.p_2) B(-\alpha^{'}s,-\alpha^{'}u)-e_2.e_4e_3.p_1 B(1-\alpha^{'}u,-\alpha^{'}s)+\nonumber \\
& &(e_3.e_4e_2.p_3-2\alpha^{'}(e_4.p_1e_3.p_1e_2.p_3+e_4.p_2e_3.p_2e_2.p_3+e_4.p_1e_3.p_2e_2.p_4))\nonumber \\
& & B(-1-\alpha^{'}s,-\alpha^{'}u)+(e_3.e_4e_2.p_4-2\alpha^{'}(e_4.p_1e_3.p_1e_2.p_4+e_4.p_2e_3.p_2e_2.p_4+e_4.p_2e_3.p_1e_2.p_3)) \nonumber \\
& &B(1-\alpha^{'}u,-1-\alpha^{'}s)-2\alpha^{'}e_4.p_1e_3.p_2e_2.p_3
B(-1-\alpha^{'}s,-1-\alpha^{'}u)-\nonumber \\&
&2\alpha^{'}e_4.p_2e_3.p_1e_2.p_4
B(-1-\alpha^{'}s,2-\alpha^{'}u)])(sin(\frac{p_1\theta p_2-p_3\theta
  p_4}{2})\nonumber \\
& &[-e_2.e_3e_4.p_1
B(1-\alpha^{'}t,-\alpha^{'}s)+(-e_2.e_3e_4.p_2+e_2.e_4e_3.p_2)B(-\alpha^{'}s,-\alpha^{'}t)+\nonumber \\
& &e_2.e_4e_3.p_1B(-1-\alpha^{'}t,-\alpha^{'}s)+(e_3.e_4e_2.p_3-2\alpha^{'}(e_4.p_1e_3.p_1e_2.p_3+e_4.p_2e_3.p_2e_2.p_3+\nonumber \\
&
&e_4.p_1e_3.p_2e_2.p_4))B(-1-\alpha^{'}s,1-\alpha^{'}t)+(e_3.e_4e_2.p_4-2\alpha^{'}(e_4.p_1e_3.p_1e_2.p_4+\nonumber \\
& &e_4.p_2e_3.p_2e_2.p_4+e_4.p_2e_3.p_1e_2.p_3))B(-\alpha^{'}t,-1-\alpha^{'}s)-2\alpha^{'}e_4.p_1e_3.p_2e_2.p_3 B(-1-\alpha^{'}s,2-\alpha^{'}t)-\nonumber \\
& &2\alpha^{'}e_4.p_2e_3.p_1e_2.p_4
B(-1-\alpha^{'}s,-1-\alpha^{'}t)])(sin(\frac{p_1\theta p_4+p_2\theta
  p_3}{2})[-e_2.e_3e_4.p_1 B(-1-\alpha^{'}u,1-\alpha^{'}t)+\nonumber \\
& &(e_2.e_3e_4.p_2-e_2.e_4e_3.p_2) B(-\alpha^{'}t,-\alpha^{'}u)+e_2.e_4e_3.p_1 B(1-\alpha^{'}u,-1-\alpha^{'}t)+\nonumber \\
& &(e_3.e_4e_2.p_3-2\alpha^{'}(e_4.p_1e_3.p_1e_2.p_3+e_4.p_2e_3.p_2e_2.p_3+e_4.p_1e_3.p_2e_2.p_4)) B(1-\alpha^{'}t,-\alpha^{'}u)+\nonumber \\
&
&(-e_3.e_4e_2.p_4+2\alpha^{'}(e_4.p_1e_3.p_1e_2.p_4+e_4.p_2e_3.p_2e_2.p_4+e_4.p_2e_3.p_1e_2.p_3))
B(1-\alpha^{'}u,-\alpha^{'}t)+\nonumber \\
& &2\alpha^{'}e_4.p_1e_3.p_2e_2.p_3 B(2-\alpha^{'}t,-1-\alpha^{'}u)-2\alpha^{'}e_4.p_2e_3.p_1e_2.p_4 B(-1-\alpha^{'}t,2-\alpha^{'}u)])
\end{eqnarray}
By going through different noncommutative amplitudes in this section we conclude that '$\sin$' factor
arises only when we have odd number of photons, which implies that in
the corresponding commutative theory the amplitude vanishes.  
\section{Conclusions}
In this study  we have demonstrated explicitly that in the
$\alpha^{'}\rightarrow 0$ limit noncommutative 
string  theory reduces to the noncommutative gauge theory and in the
said limit  no  corrections to the noncommutative gauge
theoretic action is observed (due to the absence of
massive string modes), and this is in agreement with \cite{sw}. In particular we have shown  the form of the kinematic
factor (K) in the case of the 4-point noncommutative amplitude
involving gauge bosons  to be of the same form  as that  in the
corresponding commutative theory \cite{js}  with the desired
properties,  $(1)$ Absence of tachyons, and this is in
agreement with the fact that we are dealing with a super-symmetric
theory. $(2)$ Satisfy the on-shell gauge invariance. $(3)$ It is
cyclic in the external photons.   
We also   gets sin/cos factors in the noncommutative scattering
amplitudes depending
on the number of external photons. From the study of  the
noncommutative scattering  amplitudes we conclude that the phases
arises in each sector depends on 
the ordering of the vertex operators, also the phases of the amplitudes in each sector  are not derivable
by starting from any one of the amplitude, which we can say that the total  
amplitude in noncommutative string theory is not symmetric in the
Mandelstam variables, even though the kinematic factor is symmetric, but in the total noncommutative amplitude
the presence of  sin/cos factor will be determined by the
coefficient of $e^{ik.X}$ in the vertex operator; particularly the 3-point
noncommutative amplitude of photons and tachyons leads to the appearances of  phases  in the
total noncommutative amplitude as 'sin' and 'cos' times the kinematic factors
respectively, and the case of 4-point noncommutative amplitude, we get 
the 'sin' or 'cos' factors times the kinematic factor in the
noncommutative amplitude 
depending on the presence of odd or even number external photons
respectively.  Moreover, we have shown that the couplings of
noncommutative string theory $G_s$ and the noncommutative gauge theory
$g_{YM}$ are related as $g_{YM}=\frac{G_0}{\sqrt {2\alpha^{'}}}$,  $G_0$ 
is related to $G_s$ as mentioned earlier, and the   form of the couplings
matches with \cite{sw}. This is of  the same form as that of the 
commutative theory.

\section{Acknowledgement}
I wish to thank  J. Maharana  for useful discussions, and S. Mukherji
for a critical reading of the manuscript.

%%%%%%%%%%%%%%%%%%%%%%%%%%%%%%%%%%%%%%%%%%%%%%%%%%%%%%%%%%%%


\vspace{.7in}
\begin{center}
{\bf References}
\end{center} 
\begin{enumerate}





\bibitem{sw} N. Seiberg, E. Witten, String theory and Noncommutative
  Geometry, JHEP {\bf 09}(1999)032, hep-th/9908142.
\bibitem{vs} V. Schomerus, JHEP {\bf 9906} (1999) 030, hep-th/9903205
\bibitem{ch} C-S. Chu, P-M.Ho, Noncommutative open strings and D-branes,
Nucl.Phys.{\bf B550} 151, (1999), hep-th/9812219
\bibitem{sst} N. Seiberg, L. Susskind and N. Toumbas, Strings in background 
electric field, space/time noncommutativity and a new noncritical
string theory, JHEP {\bf 06}(2000) 021, hep-th/0005040
\bibitem{gmms} R. Gopakumar, S. Minwalla, J. Maldacena, and A. Strominger, hep-th/0005048.
\bibitem{ps} B. Pioline and A. Schwarz, JHEP {\bf 08}(1999)021, hep-th/9908019,
M.M.Sheikh-Jabbari, hep-th/9911203, R. Gopakumar, J. Maldacena ,
S. Minwalla and A. Strominger, S-duality and noncommutative gauge
theory, JHEP {\bf 06}(2000)036, hep-th/0005048, J. Maharana and S.S. Pal,
Noncommutative  open string, D-brane and  Duality,
Phys.Lett.{\bf B488}:410-416, 2000, hep-th/0005113.
\bibitem{km} I.Klebanov, J. Maldacena, (1+1)-Dimensional NCOS and its U(N) Gauge theory Dual, hep-th/0006085
\bibitem{gm} J. Gomis, T. Mehen, Space-time noncommutative field theories and unitarity, hep-th/0005129.
\bibitem{agm} O. Aharony, J. Gomis and T. Mehen, On theories with light-like noncommutativity, hep-th/0006236

\bibitem{tf}Thomas Filk, Phy.Lett.{\bf B376}, 53-58, (1996)
for noncommutative complex scalar field theory II and references therein.
\bibitem{mrs} S. Minwalla, M.V. Raamsdonk and N. Seiberg,
  hep-th/9912072, M.V. Raamsdonk and N. Seiberg, hep-th/0002186.
\bibitem{jp} J. Polchinski, String theory, {\em Cambridge University Press}, Cambridge, 1998.
\bibitem{js} J. Schawrz, Physics Reports, 1982, A. Hasimoto and
I. Klebanov, Decay of excited D-branes, hep-th/9604065, M.R. Garousi and
R.C. Myers, Superstring scattering from D-branes, hep-th/9603194
\bibitem{fms} D. Friedan, E. Martinec and S. Shenker, Nucl.Phys.{\bf B271}, 93, (1986)

\end{enumerate}
\end{document}